\documentclass[12pt]{article}                     % onecolumn (standard format)

\usepackage{graphicx}
\usepackage{epsfig}
\usepackage{amsmath}
\usepackage{mathrsfs}
\usepackage{amssymb}
\usepackage{amsthm}
\usepackage[authoryear]{natbib}
\usepackage{algorithm,algorithmic}
\usepackage{url}
\usepackage[margin=1in,paperwidth=8.5in,paperheight=11in]{geometry}
\usepackage{rotating}
\usepackage{setspace} 
\usepackage{stmaryrd}
\usepackage{stackrel}
\usepackage[Bjornstrup]{fncychap}
\usepackage{hyperref}
\usepackage{listings}
\usepackage{dtklogos}

\makeatletter
\newcommand\code{\bgroup\@makeother\_\@makeother\~\@makeother\$\@makeother\^\@codex}
\def\@codex#1{{\normalfont\ttfamily\hyphenchar\font=-1 #1}\egroup}

\newcommand\proglang{\bgroup\@makeother\_\@makeother\~\@makeother\$\@makeother\^\@codex}
\def\@codex#1{{\normalfont\ttfamily\hyphenchar\font=-1 #1}\egroup}

\newcommand\pkg{\bgroup\@makeother\_\@makeother\~\@makeother\$\@makeother\^\@codex}
\def\@codex#1{{\normalfont\ttfamily\hyphenchar\font=-1 #1}\egroup}
\makeatother

\newcommand{\cellof}[1]{{\mathcal{G}[#1]}}

\renewcommand{\P}{\mathbb{P}}

\newcommand{\N}{\mathrm{N}}

\newcommand{\rmd}{\mathrm{d}}

\newcommand{\E}{\mathbb{E}}

\newcommand{\cov}{\mathrm{cov}}

\newcommand{\indic}{\mathbb{I}}

\newtheoremstyle{mytheoremstyle} % name
    {\topsep}                    % Space above
    {\topsep}                    % Space below
    {\itshape}                   % Body font
    {}                           % Indent amount
    {\scshape}                   % Theorem head font
    {.}                          % Punctuation after theorem head
    {.5em}                       % Space after theorem head
    {}  % Theorem head spec (can be left empty, meaning ‘normal’)
    
\newtheoremstyle{mydefinitionstyle} % name
    {\topsep}                    % Space above
    {\topsep}                    % Space below
    {\sf}                   % Body font
    {}                           % Indent amount
    {\scshape}                   % Theorem head font
    {.}                          % Punctuation after theorem head
    {.5em}                       % Space after theorem head
    {}  % Theorem head spec (can be left empty, meaning ‘normal’)    

\theoremstyle{mytheoremstyle}

\newlength{\descwid}
\begin{document}

\title{Spatial Modelling of Emergency Service Response Times}
\author{Benjamin M. Taylor}
\maketitle

\begin{abstract}
   This article concerns the statistical modelling of emergency service response times. We apply advanced methods from spatial survival analysis to deliver inference for data collected by the London Fire Brigade on response times to reported dwelling fires. Existing approaches to the analysis of these data have been mainly descriptive; we describe and demonstrate the advantages of a more sophisticated approach. Our final parametric proportional hazards model includes harmonic regression terms to describe how response time varies with time-of-day and shared spatially correlated frailties on an auxiliary grid for computational efficiency.
   
   We investigate the short-term impact of fire station closures in 2014. Whilst the London Fire Brigade are working hard to keep response times down, our findings suggest there is a limit to what can be achieved logistically: the present article identifies areas around the now closed Belsize, Downham, Kingsland, Knightsbridge, Silvertown, Southwark, Wesminster and Woolwich fire stations in which there should perhaps be some concern as to the provision of fire services. 
\end{abstract}

\section{Introduction}

The thought of a fire in the home terrifies most people. In 2013-2014 UK fire and rescue services attended over half a million calls; there were a total of 322 fire-related deaths and 9748 non-fatal casualties due to fires of which 80\% occurred in dwellings. There were 39600 dwelling fires in the UK in 2013-2014, with most occurring between the hours of 8 and 9pm at night and with misuse of equipment/appliances being the leading cause of around 1/3 these incidents \citep{dclg2014}. In London, 70\% of fire-related deaths have been attributed to reporting delays, but crucial in saving lives is the efficient response of emergency services to calls to 999/112 \citep{lfb2013b}.

The choice of where to locate emergency service stations (police, fire, ambulance) in cities has a direct impact on possible response times. From an academic perspective, this can be treated as an optimisation problem: how do we balance the need to respond quickly to emergency situations given a finite resource allocation \citep{toregas1971,kolesar1973}? Whilst such approaches are very useful in helping to decide where might be best to build emergency service stations in the first place, the urban environment is constantly changing and there is therefore a need to continually monitor and improve information on response times in order to ensure safety standards are maintained. One aspect of this is the profiling and mapping of high risk groups, which has been undertaken in a limited way in the UK, one example being a pair of studies in Merseyside \citep{higgins2013,higgins2014}. Another aspect is the study of response times to emergency calls, the subject of the present article.

The analysis of emergency response times has received a modest amount of attention in the literature. \cite{scott1978} is one exception to this, these authors sought to form a mathematical model for ambulance response times in Houston. Other statistical approaches have focussed on predicting demand for emergency services such as \cite{matteson2011}, \cite{vile2012} and \cite{zhou2014}. Recent concern over the UK Government's cuts to public services and their potential impact on the ability of fire services to maintain safety standards has resulted in a resurgence of interest, albeit primarily from the media and opposition parties \citep{lfb2013, lfb2013a,odi2013,bannister2014,caven2014,foley2014,johnson2014,labwm2014,read2014,mccartney2015}. The year 2014 saw 10 fire stations close: Belsize, Bow, Clerkenwell, Downham, Kingsland, Knightbridge, Silvertown, Southwark, Westminster and Woolwich.

In the present article our goal is to form a model for response times with the aim of providing emergency services with probabilistic information on where \emph{in space} response times could be improved; clearly, faster response to fires in the home means more saved lives. Whilst in this article we focus on urban fires, it is worth noting the marked difference in response times to fires in urban compared with rural areas \citep{claridge2013,torney2013}. In rural areas response times are usually longer and the results presented in the present article corroborate these findings: response times in the outskirts of the city, where there are fewer fire stations and there is generally more open space, are typically longer than near the centre of the city.

\begin{figure}[htbp]
    \centering
    \includegraphics[width=0.5\textwidth]{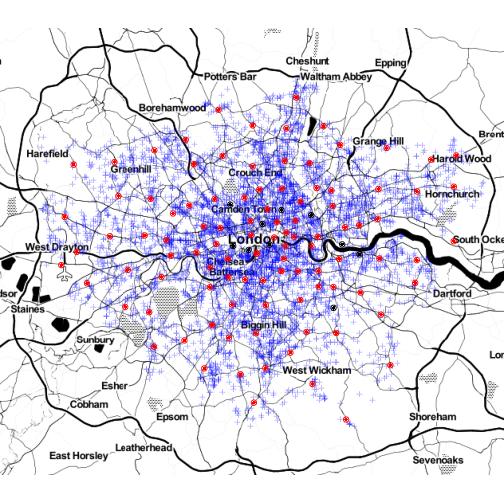}
    \caption{\label{fig:data2014} Locations of the 5769 dwelling fires in 2014 (blue crosses) and the LFB stations (red dots are open stations, black dots are closed stations).}
\end{figure}

The London Fire Brigade (LFB) is one of the largest fire and rescue services in the world; they collect and analyse substantial amounts of data on incident response times \citep{lfb2014}. In addition to presenting tables of average response time by ward, there has also been some investment in the mapping of response times at this level of aggregation \citep{lfb2013c,lfb2014a} and in preparation for the proposed closures of 2014, an assessment of the potential impact of the closures was also carried out \citep{odi2013}. Whilst these analytical efforts are to be highly commended, they could be further improved by the use of formal statistical models.

The main purpose of this article is to demonstrate ways in which the modelling-oriented approach could help to improve in the description and presentation emergency response time data and consequently inform city planners in their decision-making processes. To achieve this aim, we apply recently-developed techniques in survival analysis specifically designed for the modelling of large spatially referenced time-to-event data like the LFB response times \citep{taylor2015a}.

The remainder of this article is organised as follows. In Section \ref{sect:datamodel} we introduce some basic concepts in spatial survival analysis and give details of the modelling approach proposed; in Section \ref{sect:analysis} we present results from the analysis of the LFB data; and in Section \ref{sect:discussion}, the article concludes with a discussion.

\section{Data and Model}
\label{sect:datamodel}

The statistical analysis of time-to-event data is the realm of survival analysis \citep{cox1984,klein2013}. Most often, survival methods are applied in clinical studies assessing the potential effect of treatments or exposures on the survival time of patients. Due to patients dropping out and the fact that studies are finite in duration, survival data are typically `censored', which means the event of interest was not necessarily observed for all individuals. Survival methods handle the time to observed and censored events in a formal way.

Let $T$ be a random variable, denoting the time after the call to 999/112 that the first fire engine arrives at the scene of a dwelling fire. We will shortly introduce a model for the hazard function defined formally by, 
\begin{equation*}
    h(t) = \lim_{\Delta t\rightarrow 0}\left\{\frac{\P(t\leq T \leq t+\Delta t|T\geq t)}{\Delta t}\right\}.
\end{equation*}
For any time $t$, the interpretation of $h(t)$ in this case is as the instantaneous arrival rate of the first engine at the scene of the fire conditional on the engine having not arrived before time $t$. The survival function, which is of particular interest in our current context can be derived from the hazard function,
\begin{equation*}
    S(t) = \P(T>t) = \exp\left\{-\int_0^t h(s)\rmd s\right\},
\end{equation*}
and represents the probability that the engine will arrive on the scene after time $t$.

\begin{figure}[htbp]
    \centering
    \includegraphics[width=0.5\textwidth]{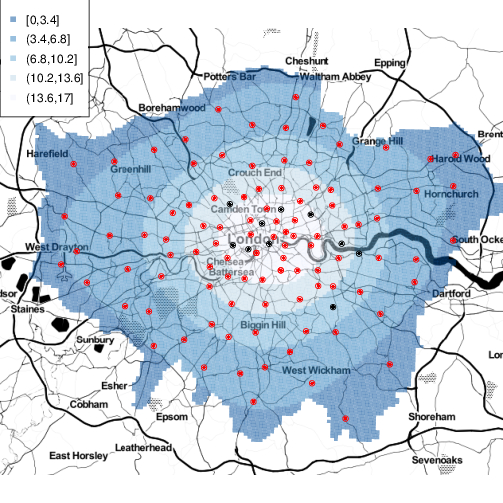}
    \caption{\label{fig:stationintensity} Smoothed intensity of fire stations. The red points are open fire stations, the black points are fire stations that closed in 2014.}
\end{figure}

Our ideal proportional hazards spatial survival model for the response times postulates the following form for the hazard function for the $i$th call:
\begin{equation}\label{eqn:hazfun}
    h(t;\beta,Y) = h_0(t)\exp\{X_i\beta + Y_i\},
\end{equation}
where $X_i$ are covariates associated with the $i$th call, $\beta$ are parameters, $h_0$ is the baseline hazard function (see below) and $Y_i$ is the value a spatially continuous Gaussian process, $Y$, at the location of the $i$th call. We assume that the Gaussian process $Y$ has associated covariance function,
\begin{equation*}
    \cov(Y_i,Y_j) = \sigma^2\rho(d_{ij};\phi)
\end{equation*}
where the parameter $\sigma^2$ is the unconditional variance of the process at any point and $\rho$ describes the interdependence of the points $i$ and $j$ at distance $d_{ij}$ apart. 

In this model $h_0$ describes the part of the hazard function that is common to all individuals and the remaining term, $\exp\{X_i\beta + Y_i\}$, describes the relative risk for the $i$th call. The relative risk splits into two parts: the first, $\exp\{X_i\beta\}$, is the part of the risk that we can explain by the available covariates; and the second, $\exp\{Y_i\}$ is the unexplained risk. With regards the latter, we choose $\E[Y]=-\sigma^2/2$, so that $\E[\exp\{Y_i\}]=1$. When survival models are used to measure time to death, a high hazard is regarded as bad, since it means there is a high chance of an individual experiencing the event. In the present context however, we need to adopt the opposite meaning in which a low hazard is interpreted as bad: at time $t$ if no engine has yet arrived and $h(t)$ is low, the chance of one arriving in the immediate future is low.

\begin{figure}[htbp]
    \centering
    \includegraphics[width=\textwidth]{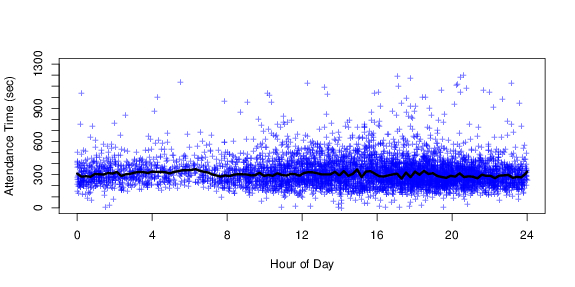}
    \caption{\label{fig:attendancebyhour2014} First pump attendance time by hour of day, the black line is a lowess smoother.}
\end{figure}

The two main options for modelling the baseline hazard are: (i) to assume a parametric form for $h_0$; or (ii) to leave $h_0$ unspecified, which results in a semiparametric model. Whilst the main advantage of the semiparametric approach is flexibility, in this article we opt for the former of these modelling paradigms because we are interested in probabilistic prediction. We considered two different parametric models for the hazard function for these data. The first is a simple Weibull model where,
\begin{equation}
   h_0(t;\alpha,\lambda) = \alpha\lambda t^{\alpha-1},\qquad \alpha,\lambda>0, \label{eqn:weibull}
\end{equation}
so that the baseline cumulative hazard takes the form:
\begin{equation*}
   H_0(t;\alpha,\lambda) = \lambda t^\alpha;\nonumber.
\end{equation*}
In the second parametric model we mimic the flexibility of a semiparametric approach by modelling the baseline hazard using B-splines as in \cite{rosenburg1995}, setting:
\begin{equation}\label{eqn:bspline}
    h_0(t;\omega) = \sum_{i=1}^p\exp\{\omega_i\}B_i^{(d)}(t),
\end{equation}
where $\omega_1,\ldots,\omega_d$ are parameters to be estimated and $B_i^{(d)}(t)$ is a B-spline basis function, a piecewise positive polynomial of degree $d$, see \cite{younes1997} for details on how to construct these. Being piecewise polynomial, the baseline cumulative hazard function $H_0(t)=\int_0^t h_0(s)\rmd s$, required in likelihood computation, is trivial to compute provided we store the (piecewise) coefficients of the integrated basis functions:
\begin{equation*}
    H_0(t;\omega) = \sum_{i=1}^p\exp\{\omega_i\}\int_0^tB_i^{(d)}(s)\rmd s,
\end{equation*}

\begin{table}[htbp]
    \centering
    \begin{tabular}{r|rrr}
          & False Alarm & Fire & Special Service \\ \hline\hline

        Aircraft & 54 & 13 & 198 \\
        Boat & 36 & 40 & 97 \\
        Dwelling & 61680 & 18769 & 60367 \\
        Non Residential & 62281 & 6460 & 8197 \\
        Other Residential & 16699 & 1363 & 1937 \\
        Outdoor & 6079 & 13845 & 6554 \\
        Outdoor Structure & 2452 & 15824 & 1278 \\
        Rail Vehicle & 68 & 47 & 156 \\
        Road Vehicle & 3410 & 6089 & 14865 \\
    \end{tabular}
    \caption{\label{tab:incgrpcat} Number of call-outs by incident group and property category between 2012 and 2014 inclusive.}
\end{table}

Our final model is a slight modification of (\ref{eqn:hazfun}) introduced in \cite{taylor2015a}; this modification concerns the frailties $Y_i$. Rather than assume these are individual-specific we adopt a shared frailty approach, introducing an auxiliary grid of cells on which we wish to predict the response times. Our model for the hazard takes the form:
\begin{equation}\label{eqn:modhazfun}
    h(t;\omega,\beta,Y) = h_0(t;\omega)\exp\{X_i\beta + Y_\cellof{i}\},
\end{equation}
where $Y_\cellof{i}$ denotes the value of the process $Y$ in the cell containing observation $i$: we approximate the spatially continuous process $Y$ by a piecewise constant process on a fine grid. The reason for doing this is primarily computational efficiency: model (\ref{eqn:hazfun}) incurs $O(n^3)$ computational cost, where $n$ is the number of observations, whereas model (\ref{eqn:modhazfun}) is $O(n)$. With there being around 6000 dwelling fires each year, a Markov chain Monte Carlo (MCMC) algorithm used to deliver inference for model (\ref{eqn:hazfun}) would be impractical.

For a discussion of other potential models for these data, see Section \ref{sect:discussion}.

\section{An Analysis of the 2012--2014 Data}
\label{sect:analysis}

The data on LFB response times analysed in this article is available from the London Datastore \citep{lfepa2015}. All analyses were carried out in the \proglang{R} statistical software \citep{rlang2014}. In this section, we present an analysis of the LFB data from 2012 to 2014 inclusive.

\subsection{Preliminary Analyses and Model Choice}

Table \ref{tab:incgrpcat} shows the number of call-outs to fires, false alarms and special services by property type for the years 2012--2014 inclusive. We restrict our attention to analysing data from call-outs to fire events in dwellings because as mentioned above, dwellings are where most fire-related deaths occur. Figure \ref{fig:data2014} shows the locations of the 5769 dwelling fires in 2014 as well as the locations of all fire stations, including those that were closed in 2014. The pattern of the points roughly follows the distribution of the population in the city with higher concentration in the centre and with large open park areas being free of dwelling fires, for instance.

\begin{figure}[htbp]
    \centering
    \includegraphics[width=\textwidth]{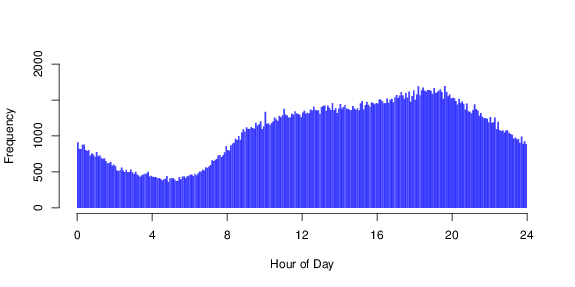}
    \caption{\label{fig:hourofday} Number of calls between 2012--2014 inclusive by hour of the day; each bar represents a 5 minute interval. The histogram includes call-outs to fires, special service and also false alarms.}
\end{figure}

It can also be seen from Figure \ref{fig:data2014} that the intensity of locations of fire stations is also more concentrated towards the centre of the city, a bivariate isotropic Gaussian smoothing of these points is shown in Figure \ref{fig:stationintensity} (the intensity was scaled by a factor of $10^8$). The bandwidth used to compute the smooth intensity was chosen using the rule of thumb method of Baddeley and Turner's \code{density.ppp} function from the \pkg{spatstat} package, resulting in a kernel standard deviation of 5737; the result is quite a smooth approximation to the intensity of the fire stations which is desirable for the use we put it to. The relationship between response time to proximity of the nearest fire station is not clear cut: it is not necessarily the closest station that will respond to a call and for this reason, we use the smoothed intensity as a covariate in our model. We expect the coefficient of this covariate to be positive, since in places where there are a greater concentration of fire stations, fires are likely to be attended more swiftly.

Whilst the plot of response time by time of day, shown in Figure \ref{fig:attendancebyhour2014} suggests no obvious trends, it is reasonable to assume that time of day does in fact influence response time due to traffic congestion and local demand on the fire service. The demand by time of day is illustrated in Figure \ref{fig:hourofday}, this shows that it is lowest at around 5:30am, highest around 7pm and generally higher in the day-time than at night. There was no obvious difference in response times comparing weekdays with weekends: the median response time in 2014 for weekdays was 299 seconds and for weekends it was 296.5; again, though there is no obvious difference between these values, it is reasonable to assume that we might expect a difference in response times due in part to differing traffic patterns on weekdays compared with weekends.

\begin{figure}[htbp]
    \centering
    \begin{minipage}{0.5\textwidth}
        \includegraphics[width=\textwidth]{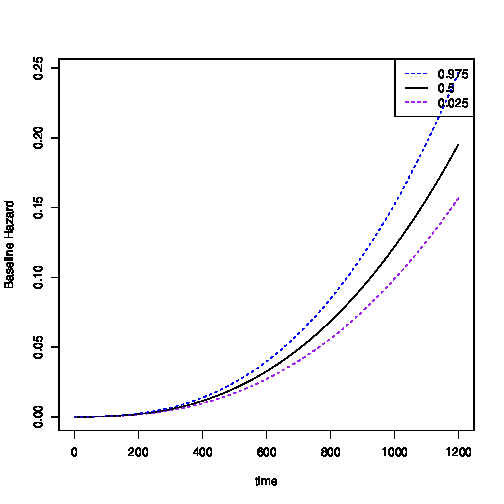}
    \end{minipage}\begin{minipage}{0.5\textwidth}
        \includegraphics[width=\textwidth]{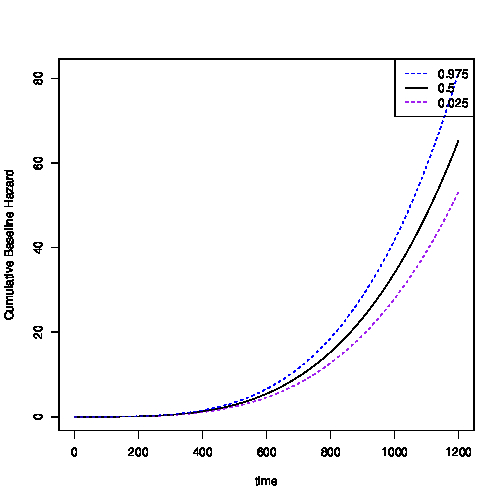}
    \end{minipage}

    \begin{minipage}{0.5\textwidth}
        \includegraphics[width=\textwidth]{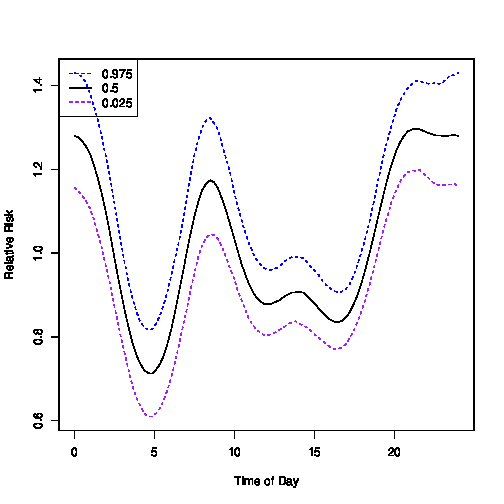}
    \end{minipage}\begin{minipage}{0.5\textwidth}
        \includegraphics[width=\textwidth]{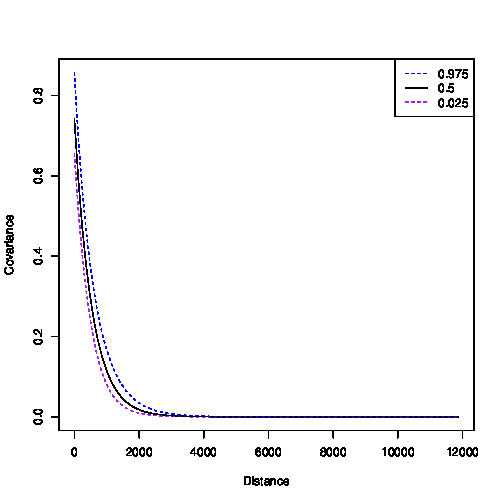}
    \end{minipage}
    \caption{\label{fig:basehaz} Top left: baseline hazard. Top right: baseline cumulative hazard. Bottom left: relative risk by time of day (compare this with Figure \ref{fig:attendancebyhour2014}, which shows no obvious trend). Bottom right: posterior covariance function. These plots are based on the results from 2014, similar plots from 2012 and 2013 are not materially different.}
\end{figure}

Our overall model for the hazard function therefore includes the intensity of fire stations and a time trend through harmonic regression terms: $\sin(2\pi kt/24)$ and $\cos(2\pi kt/24)$ with $k=1,2,3,4$ i.e. with periodicity 1 day, 1/2 day, 1/3 day and 1/4 day. The number of harmonic terms was chosen by fitting a non-spatial version of the model to the 2010 data using forward selection, terminating the process when both the $\sin$ and $\cos$ contributions were not significant at the 5\% level. We initially fitted the spatial Weibull model which included a weekday/weekend indicator variable as an additional covariate, but since this was not found to be significant, we fitted the B-spline model without this term. For the B-spline model, the baseline hazard function, $h_0$ was modelled using a piecewise cubic B-spline function of the form (\ref{eqn:bspline}) with 5 internal knots at the minimum, 0.2, 0.4 and 0.6 quantiles and at the maximum response time; with a repeated knot at each end-point the spline has a total of seven parameters. We used an exponential covariance function for the spatial random effects, that is
\begin{equation*}
\cov[Y_i,Y_j]=\sigma^2\exp\{-d_{ij}/\phi\}
\end{equation*}
where $d_{ij}$ is the distance between the centroid of the cell representing $Y_i$ and that representing $Y_j$.

\begin{table}[htbp]
    \centering
    \begin{tabular}{l|cccc}
          & Year & 50\% & 2.5\% & 97.5\% \\  \hline \hline
        intensity & 2012 & 0.101 & 7.64$\times10^{-2}$ & 0.129 \\
        weekend & 2012 & 4.83$\times10^{-2}$ & -1.55$\times10^{-2}$ & 0.113 \\
        $\alpha$ & 2012 & 3.85 & 3.76 & 3.95 \\
        $\lambda$ & 2012 & 1.05$\times10^{-10}$ & 6.05$\times10^{-11}$ & 1.89$\times10^{-10}$ \\
        $\sigma$ & 2012 & 0.879 & 0.823 & 0.94 \\
        $\phi$ & 2012 & 772 & 652 & 929 \\ \hline \hline
        intensity & 2013 & 9.82$\times10^{-2}$ & 7.85$\times10^{-2}$ & 0.118 \\
        weekend & 2013 & -4.63$\times10^{-4}$ & -6.69$\times10^{-2}$ & 6.45$\times10^{-2}$ \\
        $\alpha$ & 2013 & 3.73 & 3.64 & 3.82 \\
        $\lambda$ & 2013 & 2.18$\times10^{-10}$ & 1.21$\times10^{-10}$ & 3.76$\times10^{-10}$ \\
        $\sigma$ & 2013 & 0.859 & 0.806 & 0.916 \\
        $\phi$ & 2013 & 583 & 503 & 687 \\ \hline \hline
        intensity & 2014 & 6.4$\times10^{-2}$ & 4.52$\times10^{-2}$ & 8.2$\times10^{-2}$ \\
        weekend & 2014 & 1.64$\times10^{-2}$ & -5.79$\times10^{-2}$ & 8.52$\times10^{-2}$ \\
        $\alpha$ & 2014 & 3.58 & 3.5 & 3.69 \\
        $\lambda$ & 2014 & 6.09$\times10^{-10}$ & 3.41$\times10^{-10}$ & 1.02$\times10^{-9}$ \\
        $\sigma$ & 2014 & 0.861 & 0.809 & 0.926 \\
        $\phi$ & 2014 & 537 & 458 & 634 \\
    \end{tabular}
    \caption{\label{tab:coeffs} Table of parameter estimates from the three models fitted.}
\end{table}

\subsection{Description of Inferential Procedure}

Model (\ref{eqn:modhazfun}) was fit separately to the 2012, 2013 and 2014 datasets using the \proglang{R} package \pkg{spatsurv} \citep{taylor2014a}. This package implements a fully Bayesian adaptive MCMC algorithm, which delivers inference for the model parameters $\beta$ and $\omega$, the shared frailties $Y_\cellof{i}$ and the parameters of the process $Y$ ($\sigma$ and $\phi$). We used Gaussian priors for all parameters on an appropriately transformed scale: $\pi(\beta)\sim\N(0,100^2)$, $\log\omega\sim\N(0,10^2)$, $\log\alpha\sim\N(0,10^2)$, $\log\lambda\sim\N(0,10^2)$, $\log\sigma\sim\N(0,0.5^2)$ and $\log\phi\sim\N(\log{1000},0.5^2)$ . As per the method described in \cite{taylor2015a}, we used a $\N(0,1)$ prior for a whitened version of the spatial process, $\Gamma$ where $Y=-\sigma^2/2+\Sigma_{\sigma,\phi}^{1/2}\Gamma$, where $\Sigma_{\sigma,\phi}$ is the covariance matrix of $Y$ on the auxiliary grid; we retained the transformed samples $\{Y_\cellof{i} \}_{i=1}^{16384}$. We chose the size of cells in the computational grid to be 500m$\times$500m.

\begin{figure}[htbp]
    \centering
    \begin{minipage}{0.5\textwidth}
        \includegraphics[width=0.95\textwidth]{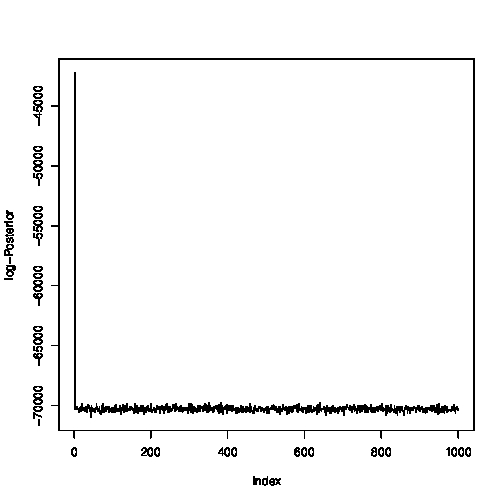}
    \end{minipage}\begin{minipage}{0.5\textwidth}
        \includegraphics[width=0.95\textwidth]{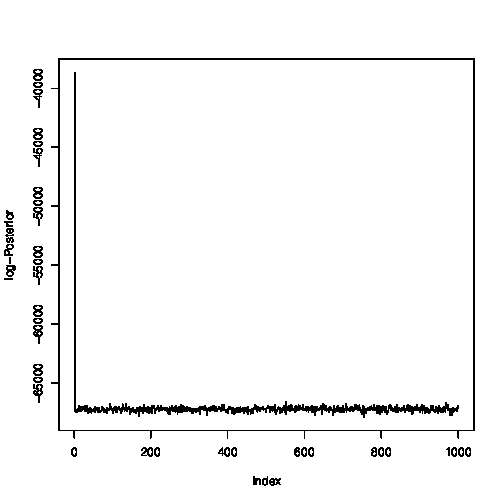}
    \end{minipage}

    \begin{minipage}{0.5\textwidth}
        \includegraphics[width=0.95\textwidth]{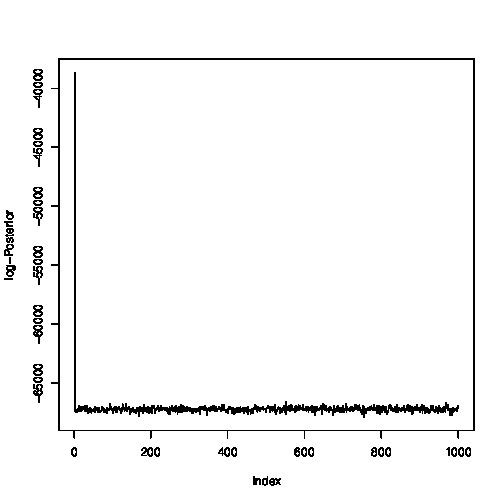}
    \end{minipage}
    \caption{\label{fig:logpost} Plot of log posterior evaluated at the initial value and over all retained iterations of the chain. This shows that initially the chain started far away from a mode, but had found one before the burnin had finished.}
\end{figure}

\begin{figure}[htbp]
	\centering

	\begin{minipage}{0.3\textwidth}
		\includegraphics[width=\textwidth]{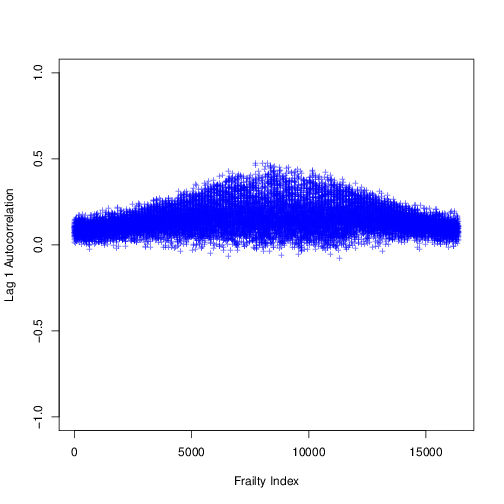}
	\end{minipage}\begin{minipage}{0.3\textwidth}
		\includegraphics[width=\textwidth]{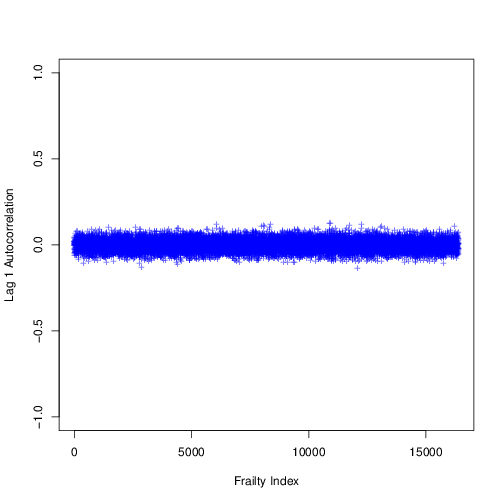}
	\end{minipage}\begin{minipage}{0.3\textwidth}
		\includegraphics[width=\textwidth]{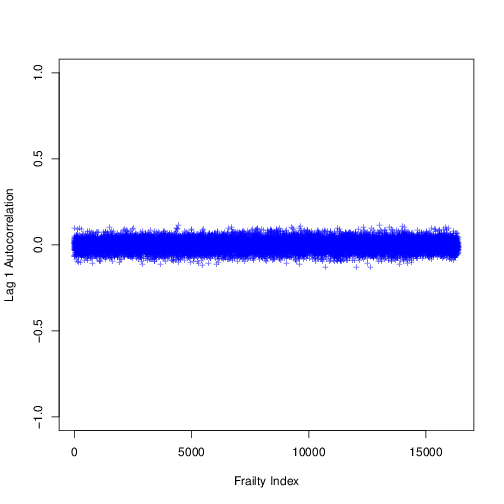}
	\end{minipage}
	
	\begin{minipage}{0.3\textwidth}
		\includegraphics[width=\textwidth]{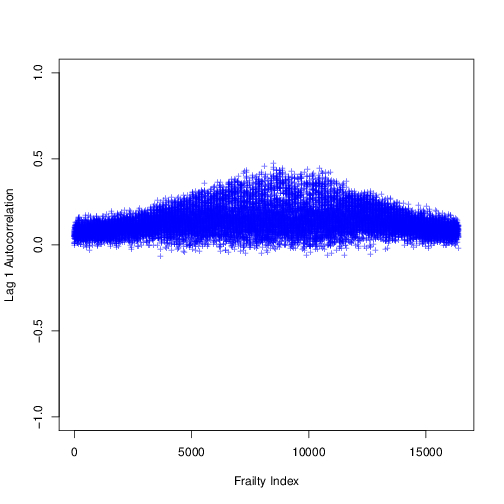}
	\end{minipage}\begin{minipage}{0.3\textwidth}
		\includegraphics[width=\textwidth]{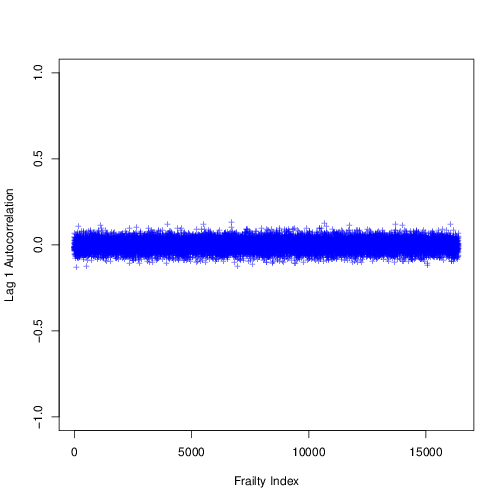}
	\end{minipage}\begin{minipage}{0.3\textwidth}
		\includegraphics[width=\textwidth]{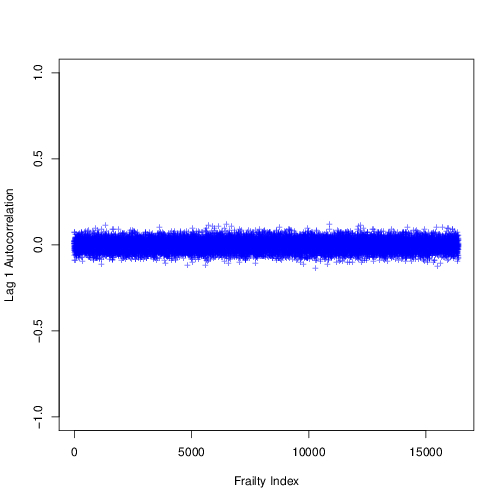}
	\end{minipage}
	
	\begin{minipage}{0.3\textwidth}
		\includegraphics[width=\textwidth]{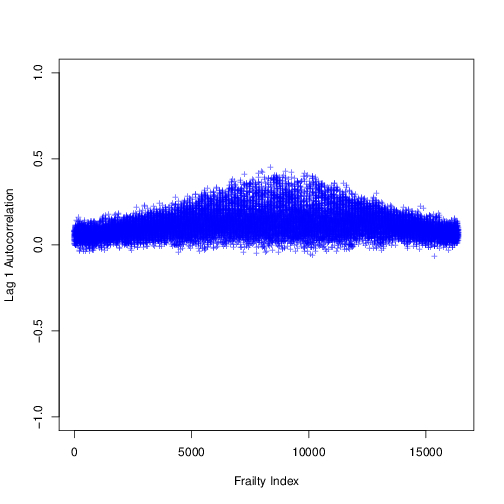}
	\end{minipage}\begin{minipage}{0.3\textwidth}
		\includegraphics[width=\textwidth]{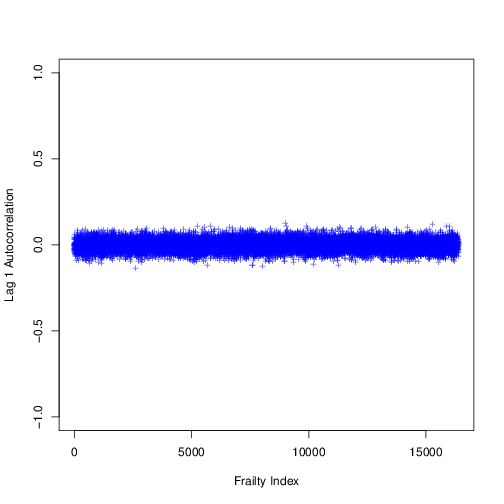}
	\end{minipage}\begin{minipage}{0.3\textwidth}
		\includegraphics[width=\textwidth]{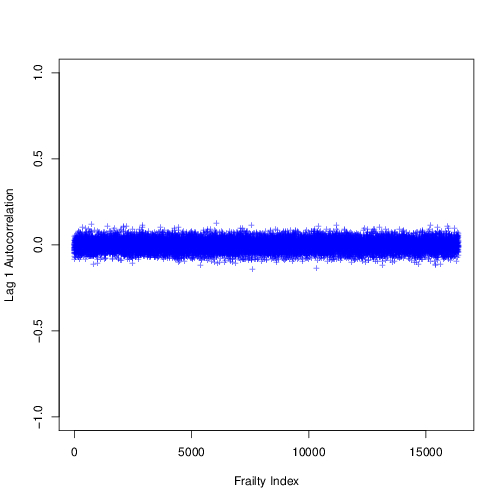}
	\end{minipage}
	
	\caption{\label{fig:frailtylag1} Autocorrelation in all sampled $Y$s. Left column: lag1, middle column: lag 5, right column: lag 10. Top row: 2014, middle row: 2013, bottom row: 2012.}

\end{figure}

For all of the Weibull models and for the 2014 and 2013 B-spline models, we ran the samplers for 500,000 iterations with a 10,000 iteration burn in and retaining every 490th sample for inference and verified convergence by examining a plot of the log posterior over the retained iterations (Figure \ref{fig:logpost}) which showed that the retained chain had left the transient phase and was at stationarity. The 2012 chain for the B-spline model required a longer burn in period and was run for 600,000 iterations with a 110,000 iteration burnin, again retaining every 490th sample. Plots of autocorrelation in the $Y$ chains (for our final Weibull model) at lags 1, 5 and 10 are shown in Figure \ref{fig:frailtylag1}, these confirm that the the chain was mixing well: the autocorrelation in all cases had dropped to a negligible amount on or before the 5th lag. The B-spline chain mixed more slowly but by lag 10 autocorrelation in the 2013 and 2014 chains was low; the 2012 chain was mixing a little more slowly. For model comparison purposes we considered these chains to be sufficiently well mixing to decide between the Weibull and B-spline models. Diagnostic plots for the other chains in our final Weibull model are available by following the links appearing in Section \ref{sect:appendices}. 

This \emph{is} a challenging sampling problem: with around 6,000 observations and for technical reasons $256\times256=65,536$ prediction points (reducing to an output grid of size $128\times128$) each chain takes around 4 days to run.

\begin{figure}[htbp]
	\centering
	\begin{minipage}{0.5\textwidth}
		\includegraphics[width=\textwidth]{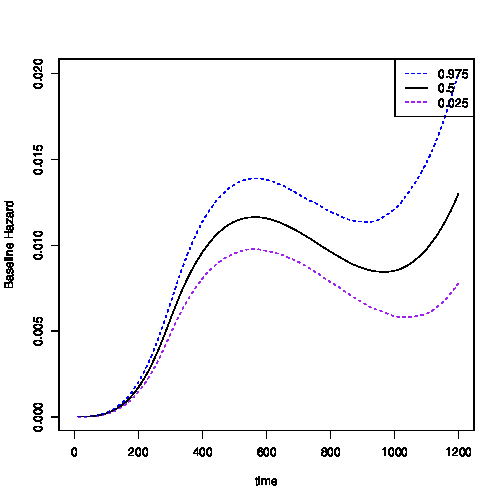}
	\end{minipage}\begin{minipage}{0.5\textwidth}
		\includegraphics[width=\textwidth]{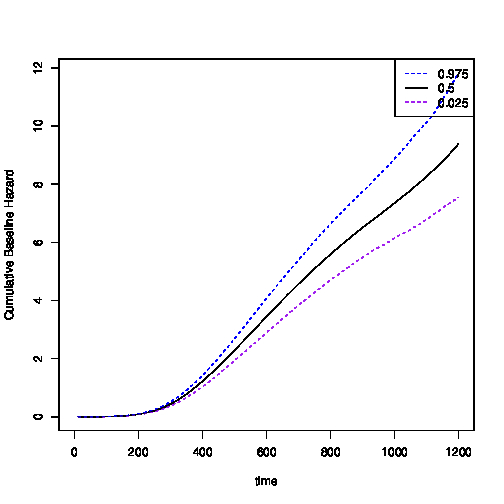}
	\end{minipage}

	\caption{\label{fig:bsplinehaz} Plot showing the estimated B-spline baseline hazard function for the 2014 data (left), and baseline cumulative hazard (right).}
\end{figure}

We chose between the Weibull and B-spline models using the DIC, shown in Table \ref{tab:DIC}. It is interesting to note that the DIC from the simpler Weibull model is lower than that for the more flexible B-spline model in each of the years considered. The parameters of the baseline hazard function are well identified by the data since every observation provides information about them; in contrast each observation provides limited information about the Gaussian process $Y$ except in the locality of the observation. The baseline hazard function for the Weibull model does not capture the shape of the hazard function well compared with the B-spline model (compare the top left plot in Figure \ref{fig:basehaz} to the left plot in Figure \ref{fig:bsplinehaz}). The lower DIC values for the Weibull model therefore suggest that the Gaussian process term in that model better explains spatial variation in residual response times compared with the B-spline model. Note that although the two baseline hazard functions look dissimilar, for times less than about 370 seconds they are in fact quite similar; 74\% of the response times in 2014 were under 370 seconds.

\begin{table}
	\centering
	\begin{tabular}{r|r|r}
		Year & B-spline & Weibull \\ \hline
		2012 & 71937.75 & 67976.35 \\
		2013 & 70454.97 & 65221.69 \\
		2014 & 66222.56 & 62177.38
	\end{tabular}
	\caption{\label{tab:DIC}Table comparing DIC values between the Weibull and B-spline models for each year.}
\end{table}

\subsection{Summary of Results}

The result of fitting this model using the is a sample from the joint posterior density of all model parameters, $\pi(\beta,\omega,Y,\eta|\text{data})$, from which we can compute expectations of quantities of interest; Figure \ref{fig:basehaz} shows several such quantities. 

Beginning with the baseline hazard, we report here the shape of the baseline hazard and cumulative hazard. The plot of the baseline hazard gives the instantaneous arrival rate conditional on there not having been an arrival so far, with the covariates and frailties in the model set to zero. Whilst the hazard function itself is more easily interpretable, it is more difficult to visualise as it varies according to location because both the fire station intensity and the frailties vary over space. The baseline hazard function is nevertheless a useful plot because it provides a global representation of the hazard, which is then scaled by the intensity of fire stations and the frailty terms in different areas of space. Under the assumed conditions, this plot shows the remarkable speed that engines arrive on the scene of a fire: the baseline hazard is quite flat for around the first 100 seconds and then starts to increase steadily. The baseline cumulative hazard (top right) can be interpreted as the expected number of arrivals of first fire engines up to a given time if those events were repeatable, the other covariates in the model being set to zero. It is difficult to see from this plot because of the scale, but under the assumed conditions, we would expect one fire engine to arrive at the scene of a fire by around 6 minutes, the London Fire Brigade target response time.

\begin{figure}[htbp]
    \centering
    \begin{minipage}{0.5\textwidth}
        \includegraphics[width=0.95\textwidth]{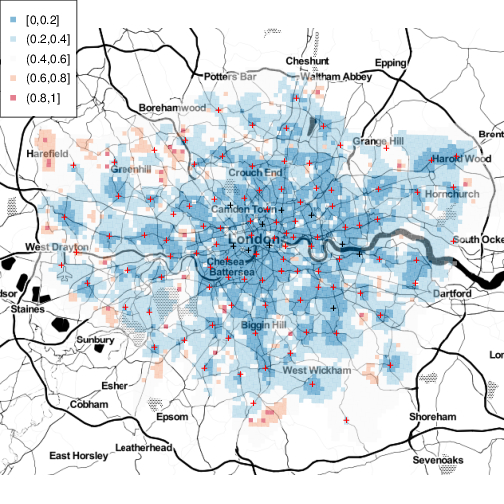}
    \end{minipage}\begin{minipage}{0.5\textwidth}
        \includegraphics[width=0.95\textwidth]{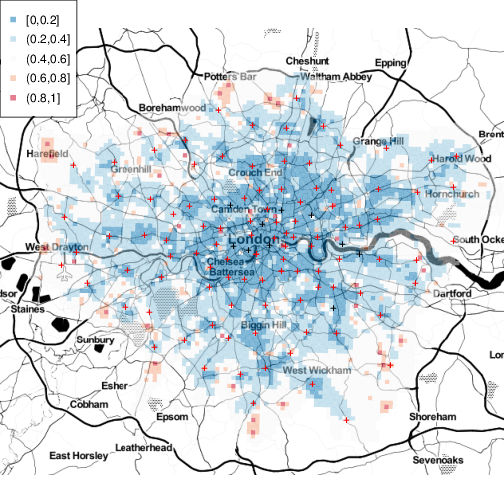}
    \end{minipage}

    \begin{minipage}{0.5\textwidth}
        \includegraphics[width=0.95\textwidth]{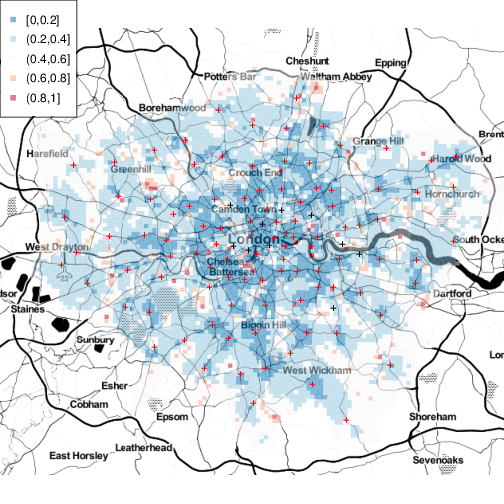}
    \end{minipage}
    \caption{\label{fig:responseExceed} Plots showing $\E[S(360)]$, the expected probability that the response time will be greater than 360 seconds. Red crosses are open stations, black crosses are stations that closed in 2014.}
\end{figure}

Whilst the baseline hazard and cumulative hazard describe global properties of the process generating these data, the arrival rate of fire engines does depend on time of day and space as will now be illustrated. The bottom left plot in Figure \ref{fig:basehaz} shows the relative risk by time of day: we do not present the coefficients of the harmonic regression terms here, as a plot is much simpler to interpret. This plot shows that there are two main times of day when services take longer than usual to arrive on the scene of a fire: between 3am and 7am and between 11am and 6pm the relative risk is significantly below 1 and reaches its lowest value of around 0.7 at 5am. It is interesting to note that this pattern was not observed in Figure \ref{fig:attendancebyhour2014}, in which arrival times appear to be independent of the time of day; adjusting for the spatial variation in risk space has therefore bought out this trend. 

The bottom right plot in Figure \ref{fig:basehaz} shows the posterior covariance function with 95\% confidence interval. This shows that spatial correlation is over quite a short range, around 0-1000 metres. Plots comparing the prior to the posterior for the parameters $\sigma$ and $\phi$ showed that these were well identified by the data (the identifiability of $\phi$ is a common problem in spatial analyses). Table \ref{tab:coeffs} gives the estimated coefficients from the Weibull model for the three years under consideration, it can be seen that the coefficients are quite similar for each year.

Using the spatial survival modelling framework, we can also illustrate answers to questions of substantive interest including (i) where in space is the London Fire Brigade's target response time of 6 minutes not being met; and (ii) what have been the effects on target response times of the 2014 fire station closures?

We answer the first question using the expected probability that the response time to a call will take longer than 6 minutes. For a response time in cell $i$ of the computational grid, this is evaluated as,
\begin{equation}
	\E[S(360)] = \frac1{1000}\sum_{i=j}^{1000} S(360;\beta^{(j)},\omega^{(j)},\eta^{(j)},Y_\cellof{i}^{(j)}),
\end{equation}
where $S(360;\beta^{(j)},\omega^{(j)},\eta^{(j)},Y_\cellof{i}^{(j)})$ is the survival function in cell $i$ evaluated for the $j$th retained sample of each parameter in the model. In computing $S$ in the above, we assumed that the fire occurred on a weekday at 8:30pm, since as stated above, most dwelling fires occur between 8 and 9pm. The resulting plot is shown in Figure \ref{fig:responseExceed}, where we have masked the computational grid cells appearing outside of the London boroughs. This plot shows that whilst there has been an improvement over the last three years in responding to reported fires in the outskirts of the city, there are some areas in the inner part of the city in 2014 where the expected probability the response will take longer than 6 minutes is slightly elevated compared with the other years -- the colour has shifted from dark blue to light blue/neutral and red in some small areas, the largest stretching from below Westminster to around Camden Town and also of note the area around Gallions Point Mariana. Some caution must be maintained in not over-interpreting these maps, as these are point estimates, they are subject to uncertainty.

\begin{figure}[htbp]
    \centering
    \begin{minipage}{0.5\textwidth}
        \includegraphics[width=0.8\textwidth]{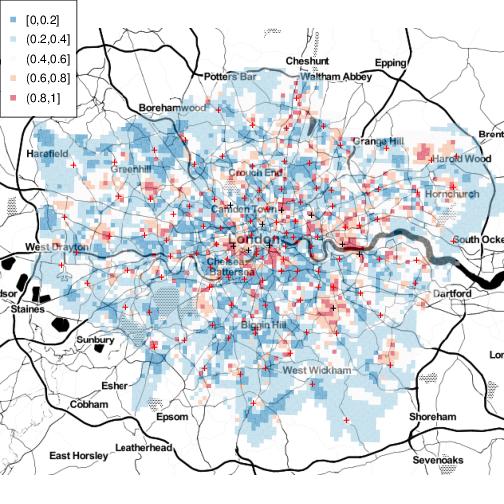}
    \end{minipage}\begin{minipage}{0.5\textwidth}
        \includegraphics[width=0.8\textwidth]{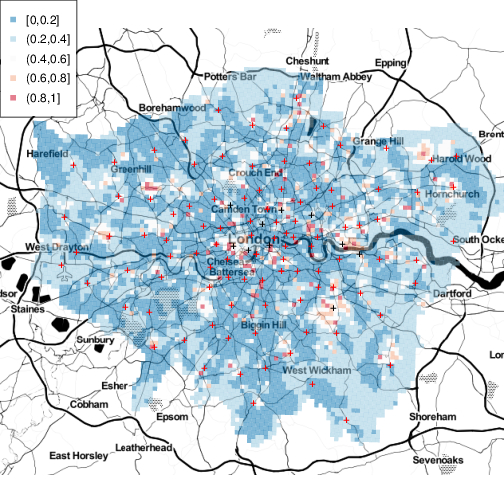}
    \end{minipage}
    
    \begin{minipage}{0.5\textwidth}
        \includegraphics[width=0.8\textwidth]{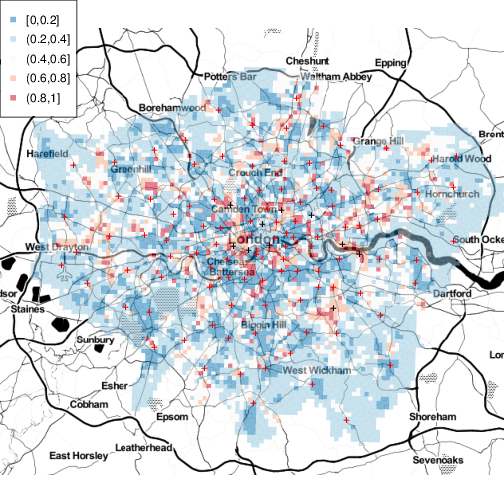}
    \end{minipage}\begin{minipage}{0.5\textwidth}
        \includegraphics[width=0.8\textwidth]{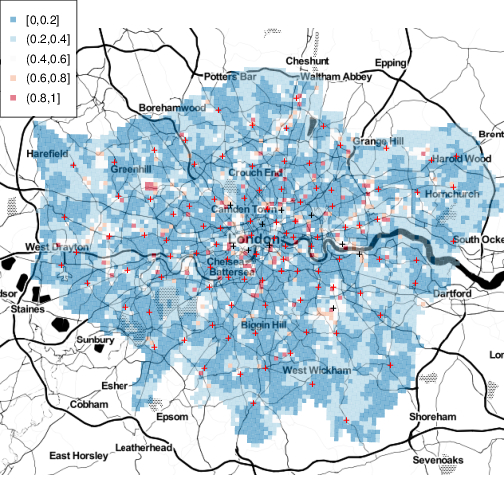}
    \end{minipage}
    
    \begin{minipage}{0.5\textwidth}
        \includegraphics[width=0.8\textwidth]{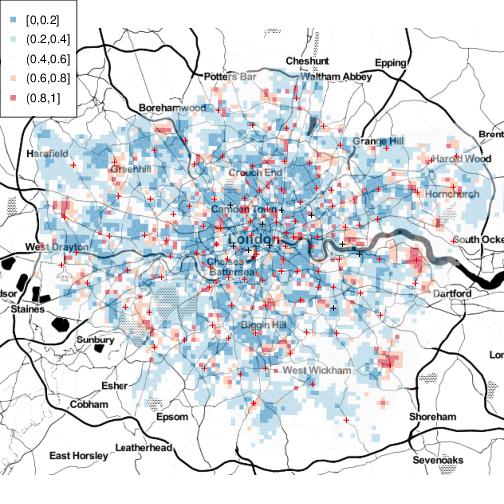}
    \end{minipage}\begin{minipage}{0.5\textwidth}
        \includegraphics[width=0.8\textwidth]{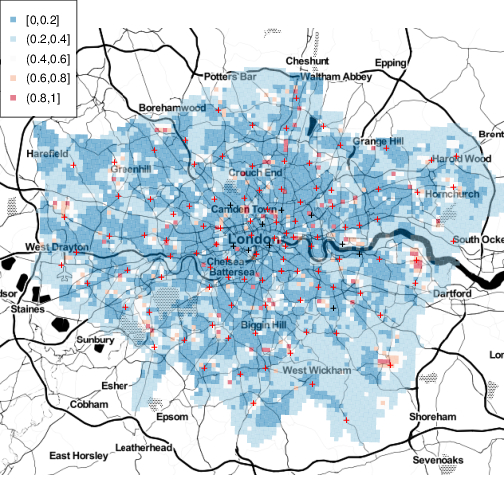}
    \end{minipage}

    \caption{\label{fig:exceed} Top row: the left plot shows $\P[S_{2014}(360)>S_{2012}(360)+0.1]$ and the right plot shows $\P[S_{2014}(360)>S_{2012}(360)+0.25]$. The middle row of plots compares 2014 with 2013 and the bottom row of plots compares 2013 with 2012; see the main text for further details. Red crosses are open stations, black crosses are stations that closed in 2014.}
\end{figure}

\begin{figure}[htbp]
    \centering

        \includegraphics[width=0.7\textwidth]{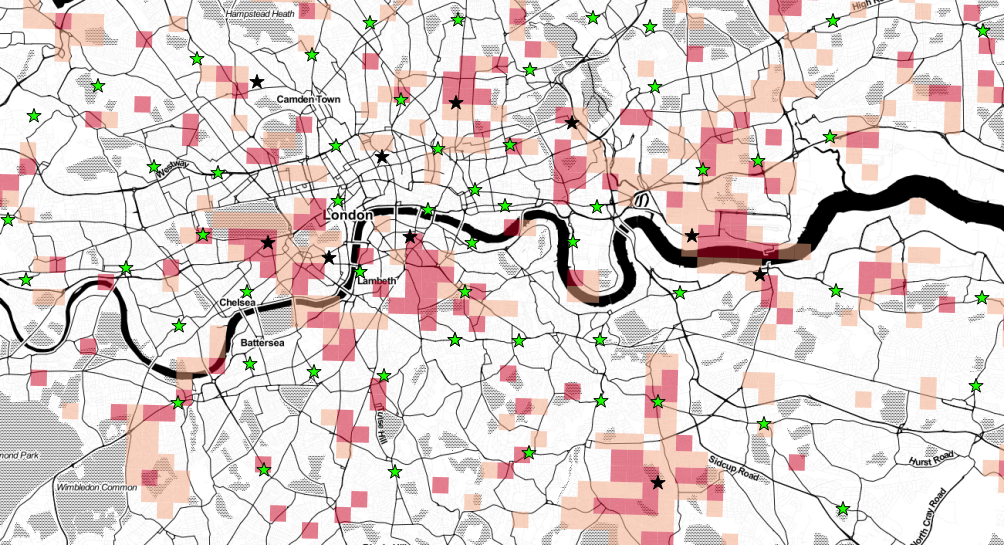}
        
        \vspace{1em}
        
        \includegraphics[width=0.7\textwidth]{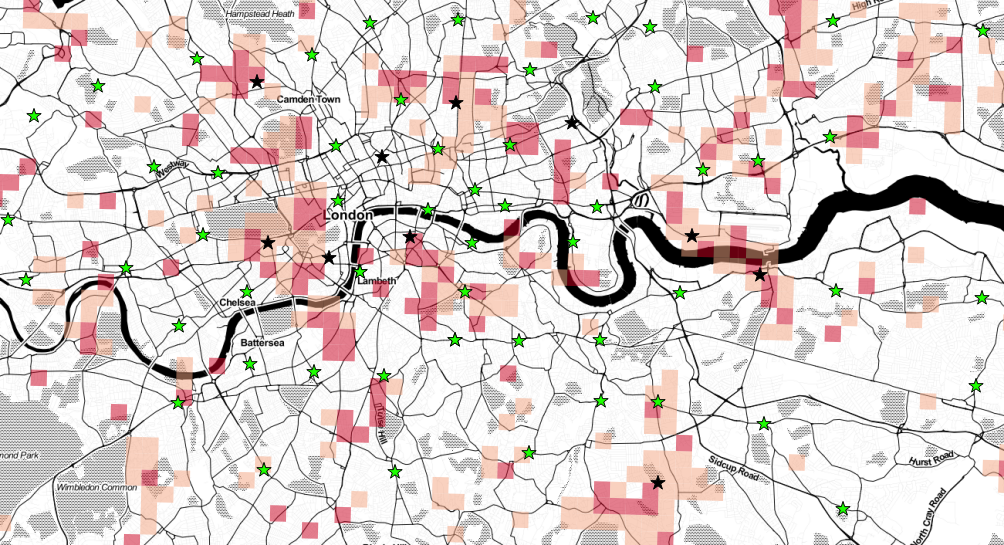}

        \vspace{1em}
        
        \includegraphics[width=0.7\textwidth]{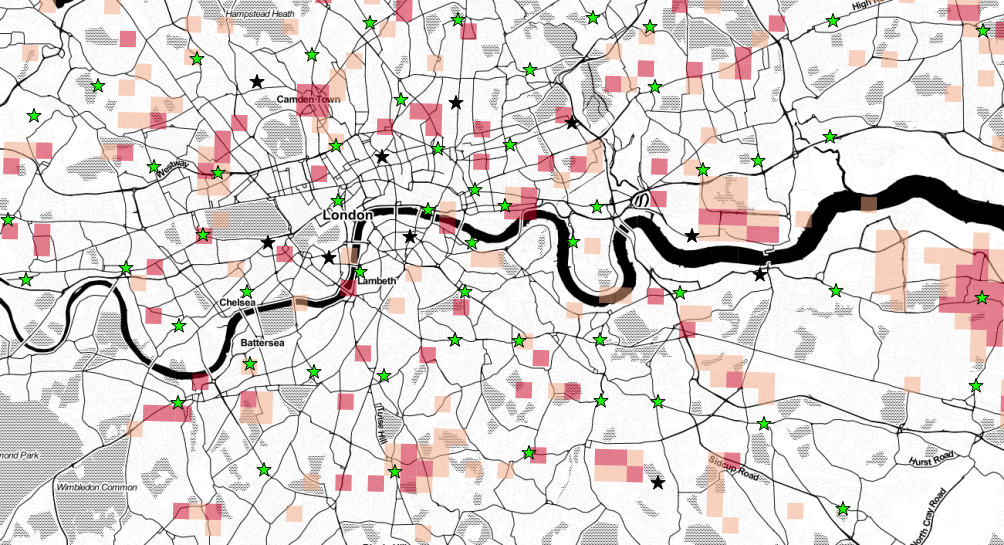}

    \caption{\label{fig:exceedzoom} Top: $\P[S_{2014}(360)>S_{2012}(360)+0.1]$. Middle Top: $\P[S_{2014}(360)>S_{2013}(360)+0.1]$. Bottom Top: $\P[S_{2013}(360)>S_{2012}(360)+0.1]$. Light red represents a probability between 0.6 and 0.8 and dark red represents a probability between 0.8 and 1. Green stars are open stations, black stars are stations that closed in 2014.}
\end{figure}

To account for this uncertainty and to make a comparison between years and thus address the second substantive question, in Figure \ref{fig:exceed}, we plotted $\P[S_{y_1}(360)>S_{y_2}(360)+c]$ for $y_1,y_2\in\{2014,2013,2012\}$ with $y_1>y_2$ and $c\in\{0.1,0.25\}$. Since we have samples from $\pi(\beta,\omega,\eta,Y|\text{data})$ in each year. In the case that $y_1=2014$, $y_2=2012$ and $c=0.1$ these probabilities were computed as:
\begin{equation}
	\frac1{1000}\sum_{j=1}^{1000}\indic[S_{2014}(360;\beta^{(j)},\omega^{(j)},\eta^{(j)},Y_\cellof{i}^{(j)})>S_{2012}(360;\beta^{(p(j))},\omega^{(p(j))},\eta^{(p(j))},Y_\cellof{i}^{(p(j))})+0.1],
\end{equation}
where $p(j)$ denotes a permuted index of the sample; in computing $S$ we again assumed that the fire occurred on a weekday at 8:30pm. The interpretation of $\P[S_{2014}(360)>S_{2012}(360)+0.1]$, for example, is the proportion of times the probability of the response time in 2014 exceeding 360 seconds is at least 0.1 bigger than the probability  of the response time in 2012 exceeding 360 seconds. Figure \ref{fig:exceed} shows these probabilities on a map of London; the top row compares 2014 with 2012, the middle row compares 2014 with 2013 and the bottom row compares 2013 with 2012. The main points of interest from these plots are the red areas, around the now closed Belsize, Downham, Kingsland, Knightsbridge, Silvertown, Southwark, Wesminster and Woolwich fire stations. Around or near to these stations there are regions where we are over 80\% confident that the probability the response time is greater than 6 minutes in 2014 were at least 0.1 bigger than in 2013 or 2012. These plots give an idea about the size and location of regions potentially affected by the closures. The area around the now closed Bow station has not been so badly affected: whilst there was an increase compared with 2012, there was not compared with 2013. Areas around the Clerkenwell station do not currently appear to have been affected, at least with respect to responses to dwelling fires.

These areas are more easily seen in Figure \ref{fig:exceedzoom}, which shows the plots for $c=0.1$ in an area around the closed stations. The important point to note here is that the spatial pattern of these probabilities comparing 2014 with 2012 and 2014 with 2013 are very similar, whereas the pattern of these probabilities comparing 2013 with 2012 is completely different. The red areas in the 2014/2012 and 2014/2013 plots are the areas that have been most affected by the closures.

\begin{figure}[htbp]
    \centering
    \begin{minipage}{0.333\textwidth}
        \includegraphics[width=0.95\textwidth]{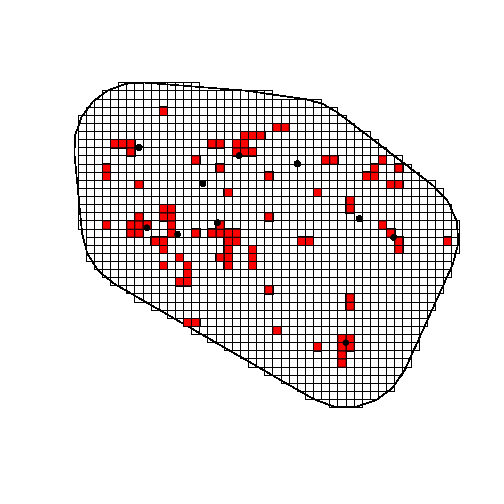}
    \end{minipage}\begin{minipage}{0.333\textwidth}
        \includegraphics[width=0.95\textwidth]{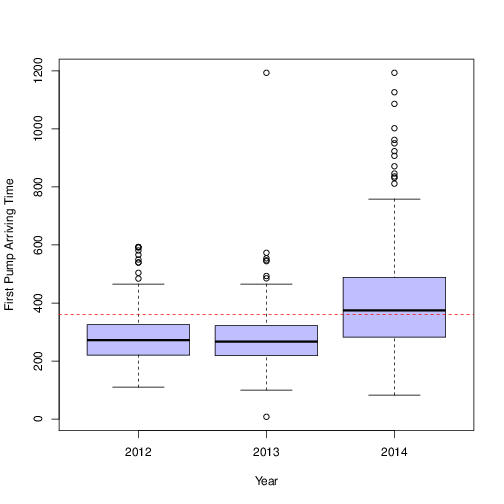}
    \end{minipage}\begin{minipage}{0.333\textwidth}
        \includegraphics[width=0.95\textwidth]{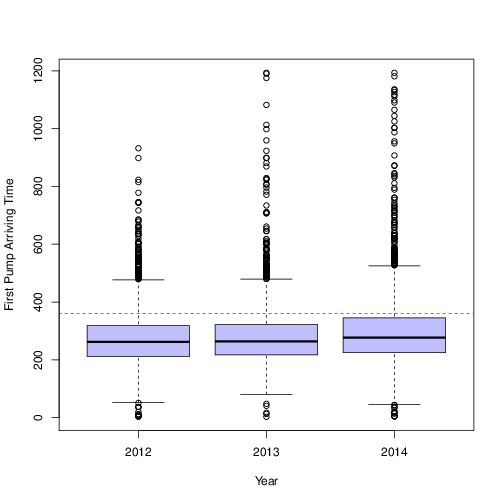}
    \end{minipage}
    \caption{\label{fig:boxplot} Box and whisker plots of attendance times in areas identified in the left plot (closed fire stations are black dots). The middle plot shows is a box and whisker plot of 705 response times in red in potentially problematic areas (see text for explanation), coloured in red in the left plot. The right hand plot is a box and whisker plot of 8276 response times in all grid cells shown. The area under consideration is the convex hull of the locations of the closed fire stations with a 4km buffer.}
\end{figure}

We used these probabilities to identify small 500$\times$500m squares near the closed fire stations that are of potential concern in terms of response time. We identified regions as those squares satisfying conditions (i) $\P[S_{2014}(360)>S_{2013}(360)+0.1]>0.7$  (ii) $\P[S_{2014}(360)>S_{2012}(360)+0.1]>0.7$ and (iii) $\P[S_{2013}(360)>S_{2012}(360)+0.1]<0.3$; i.e. areas in which response times seemed higher in 2014 compared with 2013 and 2012, but in which the probability of an increase  between 2012 and 2013 was low. We identified an area of interest around the closed fire stations by constructing the convex hull of the locations of the closed fire stations and extending it by a buffer zone of 4km. The left plot in Figure \ref{fig:boxplot} shows in red the regions meeting criteria (i) to (iii) above. The middle plot in this figure is a box and whisker plot of response times in the red regions; there were a total of 705 calls in 2012--2014 in these small areas. The right hand plot is the same but illustrates times for all small regions inside the area of interest; there were are total of 8276 of these. Calls within the red regions of potential concern therefore accounted for 9\% of all calls in the area of interest surrounding the closed stations. It can be seen from the two box and whisker plots that whilst in the region as a whole, the fire brigade appears to mainly be meeting their 6 minute target (right plot), in these small areas of potential concern near to the closed fire stations, there is a definite increase in response times in 2014; the median in these areas is above the 6 minute target.

\section{Discussion}
\label{sect:discussion}

In this article we have shown how advanced methods from spatial survival analysis can be used to model emergency service response times. We have applied these methods to the London Fire Brigade data and have illustrated the impact of the 2014 closures on response times. We have identified areas of potential concern surrounding the recently closed stations; in these areas the median response time exceeds the London Fire Brigade's target of six minutes. 

Whilst there may be simpler ways to model these data such as kriging the log-response times, the fitting of a spatial survival model is advantageous, being a `natural' model for these time-to-event data. Secondly, the hazard, cumulative hazard and survival functions are useful for describing properties of the data that are of genuine interest in this context.

The proposed inferential framework is advantageous as it provides a sample from the joint posterior of all model parameters including the spatial process $Y$ on all prediction cells. These samples can be used to deliver posterior expectations of functionals of interest, some of which have been illustrated in this article. The main drawback with the proposed method is computational cost. We estimate it would take over 5 months to run the full model in Equation (\ref{eqn:hazfun}), so the 4 days the sampler takes represents a substantial reduction in cost. Other techniques to speed up the MCMC such as using a Gaussian Markov random field to represent the spatial process on the auxiliary grid are also not as fast as the proposed method: the Fourier methods applied here scale as $O(m\log m)$ compared with $O(m^{3/2})$ for sparse matrix methods, where $m$ is the number of grid cells.

\section*{Ackowledgements}

Map tiles by Stamen Design, under CC BY 3.0. Data by OpenStreetMap, under ODbL

The London Fire Brigade data in this article are subject to the Open Government License: \url{http://www.nationalarchives.gov.uk/doc/open-government-licence/}.

\section{Appendices}
\label{sect:appendices}

\begin{itemize}
 \item 2012 diagnostic plots: \url{http://www.lancaster.ac.uk/staff/taylorb1/londonfires/2012/traceplots_2012.html} 
 \item 2013 diagnostic plots: \url{http://www.lancaster.ac.uk/staff/taylorb1/londonfires/2013/traceplots_2013.html} 
 \item 2014 diagnostic plots: \url{http://www.lancaster.ac.uk/staff/taylorb1/londonfires/2014/traceplots_2014.html} 
\end{itemize}

\newpage

\bibliographystyle{chicago}
\bibliography{bibliography}

\end{document}